\documentclass[seceq,preprint]{ptptex}

\usepackage{graphicx}
\usepackage{amsfonts}
\usepackage{amssymb}
\usepackage{latexsym}

\preprintnumber[3cm]{
OCU-PHYS-306\\AP-GR-63\\PTPTeX ver.0.8\\ \today}

\title{
Relativistic Gravitational Collapse of a Cylindrical Shell
  of Dust II
}

\subtitle{Settling Down Boundary Condition}    

\author{
$^1$Ken-ichi \textsc{Nakao}, 
$^2$Tomohiro \textsc{Harada},  
$^3$Yasunari \textsc{Kurita} \\
and \\
$^4$Yoshiyuki \textsc{Morisawa} 
}

\inst{

$^1$Department of Mathematics and Physics,~Graduate School of Science, \\
Osaka City University, Osaka 558-8585,~Japan. \\
$^2$Department of Physics,~Rikkyo University, Toshima, Tokyo~171-8501,~Japan.\\
$^3$Kanagawa Institute of Technoligy, Shimo-Ogino, Atsugi, Kanagawa 243-0292,~Japan
\\
$^4$Faculty of Liberal Arts and Science,\\
Osaka University of Economics and Law, 6-10 Gakuonji, Yao, \\
Osaka 581-8511,~Japan
}

\date{\today}

\abst{
We numerically study the dynamics of an imploding hollow cylinder composed of dust. 
Since there is no cylindrical black hole in 4-dimensional
spacetime with physically reasonable energy conditions, a collapsed dust
cylinder involves a naked singularity accompanied by its causal future, 
or a fatal singularity which terminates the history of the whole universe. 
In a previous paper, the present authors have shown that if the dust is assumed to be composed 
of collisionless particles such that these particles go through the symmetry axis 
of the cylinder, then the scalar polynomial singularity formed on the symmetry axis 
is so weak that almost all of geodesics are complete, and thus effectively 
no singularity forms by the collapse of a hollow dust cylinder. 
By contrast, in this paper, we assume that whole of 
the collapsed dust settles down on the symmetry axis by changing its equation of state. 
Obtained solutions are the straightforward 
extension of Morgan's null dust solution, in which no
gravitational radiation is emitted. 
However, in the present case with timelike dust, 
infinite amount of $C$-energy initially stored 
in the system is released through gravitational radiation. 
We also show that the gravitational waves asymptotically 
behave in a self-similar manner. 
}

\begin{document}

\maketitle

\section{Introduction}

Gravitational collapse of massive objects and formation of
spacetime singularities are one of the most prominent phenomena
predicted by general relativity.\cite{Hawking-Ellis} 
Physical quantities will blow up at the spacetime singularities 
and thus all of known theories of physics are not applicable to describe 
physical processes realized there. This means that new physics must exist 
at the spacetime singularities\cite{Harada:2004}. 
An important issue related to the attributes of the spacetime singularities is 
known as the so called cosmic censorship conjecture\cite{Penrose:1969}. 
There are two versions, the weak and strong cosmic censorship conjectures. 
The weak version states that the singularities produced by 
gravitational collapse are generically contained 
in black holes, whereas the strong version
asserts that, generically, timelike singularities do not occur. 
A few serious counterexamples for strong version 
have been found\cite{Ori:1992,Burko:1997,Dafermos:2003}, 
while the weak version is still anecdotal. Thus, hereafter, we mainly 
focus on the weak censorship. A more precise formulation of 
the weak cosmic censorship conjecture 
was given by Wald.\cite{Wald:1997}

\vskip0.3cm
\noindent
{\bf Weak cosmic censorship conjecture}: Let $\Sigma$ be a 3-manifold which, 
topologically, is the connected sum of ${\cal R}^3$ and a compact manifold. 
Let $(h_{ab}, K_{ab},\psi)$ be nonsingular, asymptotically flat initial data on $\Sigma$ 
for a solution to Einstein's equation with {\it suitable} matter (where $\psi$ denotes the 
appropriate initial data for the matter). Then, generically, the maximal Cauchy 
evolution of this data is a spacetime, $({\cal M}, g_{ab})$ which is asymptotically 
flat at future null infinity with complete ${\cal I}^{+}$.

There are two types of the counterexamples for the weak cosmic censorship 
conjecture; the first type is that effects of spacetime singularities propagating to 
infinities is so fatal that the history of the universe ceases, 
and the second type is that gentle physical effects of spacetime 
singularities propagate to infinities. In the first type, 
the causal futures of the 
spacetime singularities do not exist, and thus the spacetime 
will be globally hyperbolic. The counterexamples of this type are not the counterexamples 
for the strong cosmic censorship conjecture. By contrast, in the second type, 
the spacetime singularities must be accompanied by 
their causal futures, i.e., domains which suffer the physical 
effects of the spacetime singularities; such singularities are called 
the naked singularities. This type is also the counterexample of the strong 
cosmic censorship conjecture. 

Even though we have not yet experienced fatal influences of spacetime singularities, 
we cannot assert that the first type counterexample for the weak cosmic 
censorship does not exist in our universe. As mentioned, the spacetime singularity will 
be a signal of the violation of general relativity and implies 
occurrence of quantum gravitational phenomena. 
Thus even if fatal influences 
of singularities are predicted by general relativity, the real effects can be 
mildened by quantum effects. 

One of the most impressive example was numerically presented by Shapiro and  
Teukolsky\cite{Shapiro:1991}. They showed that gravitational collapse of 
a highly elongate mass composed of collisionless particles might be 
a counterexample for the weak cosmic censorship conjecture 
in accordance with the hoop conjecture which 
states that a black hole with horizon forms when and only 
when the mass $M$ gets compacted into a region whose circumference in every 
direction is $C\leq4\pi GM/c^{2}$.\cite{Thorne:1972}
Exactly speaking, the numerical analyses 
by Shapiro and Teukolsky did not show that the weak cosmic censorship
conjecture breaks down, 
but showed only blowing-up tendencies of physical quantities 
without a trapped surface. Anyway, in order to investigate whether counterexamples 
for the weak cosmic censorship conjecture exist, we need try to  
solve Einstein equations as a Cauchy problem 
to follow the time evolution of the 3-dimensional space like as 
the analyses by Shapiro and Teukolsky. 
However, here, it should be noted that we can know whether 
the weak cosmic censorship conjecture holds,  
only when we specify boundary conditions on the spacetime singularities, or 
in other words, the physical nature of the spacetime singularities. The 
analyses by Shapiro and Teukolsky lack this point. 

In this paper, we study the gravitational collapse of a hollow cylinder composed of 
dust. This system is not asymptotically flat and thus is out of the scope of 
the weak cosmic censorship. 
However, here we extend the weak cosmic censorship 
conjecture to the spacetimes which have 
a translational invariance in one direction and the asymptotically flat nature 
in its perpendicular directions. The case of our present interest 
is within the scope of this extended weak cosmic censorship conjecture. 

No black hole (black cylinder) forms 
by gravitational collapse in 4-dimensional spacetime, 
if physically reasonable energy conditions 
hold.\cite{Thorne:1972,Hayward:2000} Thus it is widely believed that 
the gravitational collapse of 
a dust cylinder is a counterexample for the extended 
weak cosmic censorship 
conjecture\cite{Morgan:1973,Echeverria:1993,Letelier:1994,Chiba:1996,Nolan:2002}.
However, it is not necessarily true. The present authors studied the same subject 
in paper I,\cite{NKMH:2007} and showed that by imposing one of physically reasonable 
boundary conditions at the symmetry axis of the cylinder, 
the spacetime singularity becomes very mild, and, as a result, almost all of geodesics 
are complete. The gravitational collapse of a hollow dust cylinder may not be 
a counterexample for the extended weak cosmic censorship conjecture. 
This fact will have an important meaning also for the original 
weak cosmic censorship conjecture. 
Wald required that the matter should be {\it suitable} such that, in any fixed, globally 
hyperbolic background spacetime (such as Minkowski spacetime), one always 
obtains globally non-singular solutions of the matter field equations starting from 
regular initial data. The dust does not satisfy Wald's requirement. 
However, the example of the collapsed hollow dust cylinder implies 
that Wald's requirement is too restrictive.   
It is well known that gravitational collapse of 
spherical perfect fluid can form a central shell focusing 
naked singularity\cite{Eardley:1979,Christodoulou:1984,Newman:1986,Joshi:1993,Harada:1998} 
which can be strong in Tipler's 
sense\cite{Tipler:1977,Clarke:1993}. The perfect fluid also does not satisfy the Wald's requirement. 
However this central shell focusing 
singularity is massless,\cite{Lake:1992} and hence it is non-trivial issue 
whether these examples are so serious that these should be 
regarded as counterexamples for the weak cosmic censorship conjecture. 

In paper I, we required that the dust particles  
do not stay on the symmetry axis, but pass through there. This requirement is equivalent 
to the assumption that the dust is composed of collisionless particles. 
By contrast, in this paper, we require that after dust particles arrive at 
the symmetry axis, these stay there. This is a straightforward extension of the Morgan's 
null dust solution,\cite{Morgan:1973} and it is one of our purposes to clarify the 
difference of the dust solution from the Morgan's null dust solution.  
This requirement is equivalent to the assumption that very high density 
makes interactions between particles strong. 
By this condition, a remnant of the collapsed dust cylinder remains on the symmetry axis. 

This paper is organized as follows. In $\S$2, we show the canonical 
coordinate system for the spacetime with whole cylinder symmetry and the Einstein 
equations in this coordinate system. In $\S$3, 
we briefly review the Morgan's solution which describes the collapse of a hollow 
cylinder composed of null dust. In $\S$4, we give the basic equations 
for the dust case, and discuss the boundary conditions on the metric components and 
matter variables at the spacetime singularity. Then, in $\S$5, numerical results are shown. 
In $\S$6, we give analytic solutions which asymptotically agree well with 
the numerical solutions for the metric variables. This analytic solutions 
imply that infinite energy per unit translational Killing length 
initially stored in the system is released by gravitational radiation. 
Finally, $\S$7 is devoted to summary and discussion. 

In this paper, we adopt the unit of $c=1$. We adopt the abstract index notation; 
the Latin indices $a$, $b$, $c$ denote the type of the tensor, while 
the Greek indices $\mu$, $\nu$, $\rho$, $\sigma$ mean the components 
with respect to the coordinate basis. 

\section{Basic Equations for Cylindrical System}

The spacetime with whole cylinder symmetry 
is defined by the following metric,\cite{Melvin:1964,Melvin:1965} 
\begin{equation}
ds^{2}=e^{2(\gamma-\psi)}\left(-dt^{2}+dr^{2}\right)
+R^2e^{-2\psi}d\varphi^{2}+e^{2\psi}dz^{2}, \label{eq:line}
\end{equation}
where $0\leq r<+\infty$, $0\leq\varphi<2\pi$ and 
$-\infty<z<+\infty$ constitute the cylindrical coordinate system, and $\gamma$, 
$\psi$ and $R$ are functions of $t$ and $r$. This coordinate system is called the 
canonical coordinate.  
In order that $r=0$ corresponds to symmetry axis, $R$ should vanish at
$r=0$. The coordinate variables $t$, $r$ and $z$ are all normalized so as 
to be dimensionless. 

The Einstein equations for the line element (\ref{eq:line}) are 
\begin{eqnarray}
&&\gamma'=
\left({R'}^2-{\dot R}^2\right)^{-1}
\biggl[RR'\left({\dot \psi}^{2}+{\psi'}^{2}\right)
-2R\dot{R}{\dot \psi}\psi' +R'R''-\dot{R}\dot{R}' \nonumber \\
&&~~~~-8 \pi G \sqrt{-g} \left(R'T^t{}_t+\dot{R}T^r{}_t\right)
\biggr], \label{eq:constraint-01}\\
&&\dot{\gamma}=
-\left({R'}^2-{\dot R}^2\right)^{-1}
\biggl[R\dot{R}\left({\dot \psi}^{2}+{\psi'}^{2}\right)
-2RR'{\dot \psi}\psi' +\dot{R}R''-R'\dot{R}' \nonumber \\
&&~~~~-8 \pi G \sqrt{-g} \left(\dot{R}T^t{}_t+R'T^r{}_t\right)
\biggr], \label{eq:constraint-02}\\
&&{\ddot \gamma}-\gamma''={\psi'}^{2}-{\dot \psi}^{2}
-\frac{8\pi G}{R}\sqrt{-g}T^\varphi{}_\varphi, \label{eq:gamma-evol-1}\\
&&{\ddot R}-R''=-8\pi G\sqrt{-g}\left(T^t{}_t+T^r{}_r\right), \label{eq:R-evol-0}\\
&&{\ddot \psi}+{{\dot R}\over R}{\dot \psi}-\psi''
-{R'\over R}\psi' 
=-{4\pi G\over R}\sqrt{-g}\left(T^t{}_t+T^r{}_r+T^\varphi{}_\varphi-T^z{}_z\right). 
\label{eq:psi-evol-0}
\end{eqnarray}
where $g$ is the determinant of the metric tensor, and 
a dot represents the derivative with respect to $t$, 
while a dash represents the derivative with respect to $r$. 

\section{Imploding null dust}

It is instructive to see the gravitational 
collapse of an imploding hollow cylinder composed of null 
dust, before studying the case of dust.
An exact solution was given by Morgan,\cite{Morgan:1973} and was studied by 
several authors\cite{Letelier:1994,Nolan:2002,Kurita:2006}. 
The stress-energy tensor of the null dust is given by
\begin{equation}
T^{ab}=\rho k^a k^b, \label{eq:ST-null}
\end{equation}
where $k^a$ is the vector field tangent to future-directed 
ingoing radial null geodesics, and 
$\rho$ is assumed to be non-negative function so that 
physically reasonable energy conditions hold. 
Non-trivial components of geodesic equations $k^a\nabla_ak^b=0$ are
\begin{equation}
\frac{d k_t}{du}=0~~~~{\rm and}~~~~\frac{d k_r}{du}=0, 
\end{equation}
where $u$ is the affine parameter and we have used the null 
condition $k^ak_a=0$. For imploding null dust, we have
\begin{equation}
k_\mu=(-1,-1,0,0). \label{eq:k-sol}
\end{equation}
Using the geodesic equations  $k^a\nabla_ak^b=0$, 
the equation of motion $\nabla_a T^{ab}=0$ 
becomes 
\begin{equation}
(\partial_t-\partial_r)(R\rho)=0.
\end{equation} 
The general solution of the above equation is
\begin{equation}
\rho=\frac{D(w)}{R}, \label{eq:rho-sol}
\end{equation}
where $w=t+r$ is the advanced time and $D$ is an arbitrary non-negative 
function of $w$. 

The right hand sides of Eqs.(\ref{eq:R-evol-0}) and (\ref{eq:psi-evol-0}) vanish 
identically for null dust. Everywhere finite solutions for these equations are
\begin{equation}
\psi(t,r)=\int_r^\infty\frac{f(t+x)-f(t-x)}{\sqrt{x^2-r^2}}dx,
\end{equation}
and 
\begin{equation}
R=g(t+r)-g(t-r), 
\end{equation}
where $f$ and $g$ are arbitrary functions. 
Here, in accordance with Morgan, we assume $f=0$ and $g(y)=y/2$. 
Then we have
\begin{equation}
\psi=0~~~~~{\rm and}~~~~~R=r. \label{eq:R-sol}
\end{equation}
Using the above solutions, 
Eqs. (\ref{eq:constraint-01}) and (\ref{eq:constraint-02}) lead 
\begin{eqnarray}
(\partial_t-\partial_r)\gamma&=&0, \\
(\partial_t+\partial_r)\gamma&=&8\pi GD.
\end{eqnarray}
From the above equations, we find that $\gamma$ is also a function 
of the advanced time $w$ only, and we have
\begin{equation}
\gamma=4\pi G\int_{-\infty}^wD(x)dx.
\end{equation}
We assume that the function $D(w)$ has a compact support $(w_{\rm i},w_{\rm o})$ 
so that the null dust forms a hollow cylinder. 
Due to this assumption, $\gamma$ vanishes for $w<w_{\rm i}$, and thus the symmetry axis 
$r=0$ is regular for $t\leq w_{\rm i}$, i.e., before the null dust 
reaches the symmetry axis.  For $w_{\rm i}<w<w_{\rm o}$, $\gamma$ is 
an increasing function of $w$ due to 
the non-negative nature of $D$, whereas $\gamma$ is a positive constant for 
$w\geq w_{\rm o}$. Non-vanishing $\gamma$ at $r=0$ 
implies the conically singular symmetry axis, and thus the 
symmetry axis is conically singular for $t>w_{\rm i}$. 

We can easily see from Eqs.(\ref{eq:rho-sol}) and (\ref{eq:R-sol}) 
that $\rho$ is infinite at $r=0$ if $D$ does not vanish there. 
Thus for $w_{\rm i}<t<w_{\rm o}$, the energy density $\rho$ diverges at $r=0$. 
Although all of the scalar polynomials 
composed of the Riemann tensor vanish, the components of the 
Riemann tensor with respect to a frame parallelly propagated along 
a timelike geodesic connected to the symmetry axis diverge 
for $w_{\rm i}<t<w_{\rm o}$ in the limit of $r\rightarrow0$.\cite{Letelier:1994} 
Hence the symmetry axis $r=0$ for $w_{\rm i}<t<w_{\rm o}$ is the so-called 
{\it p.p.} singularity.\cite{Hawking-Ellis} This singularity is naked and 
satisfies the strong curvature condition defined 
by Kr\`{o}lak\cite{Krolak:1987,Kurita:2006}. 
$D$ vanishes on the symmetry axis for $t \geq w_{\rm o}$, but the regularity is 
not recovered, since, as mentioned above, the symmetry axis is conically 
singular also for $t\geq w_{\rm o}$. 
This conical singularity implies that a singular line source 
remains there.\cite{Israel:1977} Thus this solution describes a 
process that the the null dust collapses into the symmetry 
axis and settles down there by changing its equation of state. 
It is well known that such a singularity is obtained in the thin limit of 
a straight cosmic string, but converse is not necessarily true. 
As Geroch and Traschen discussed, the equation 
of state for the matter condensed into this singularity cannot be 
uniquely specified.\cite{Geroch:1987}

\begin{figure}
\centerline{\rotatebox{0}{\resizebox{8cm}{!}{\includegraphics{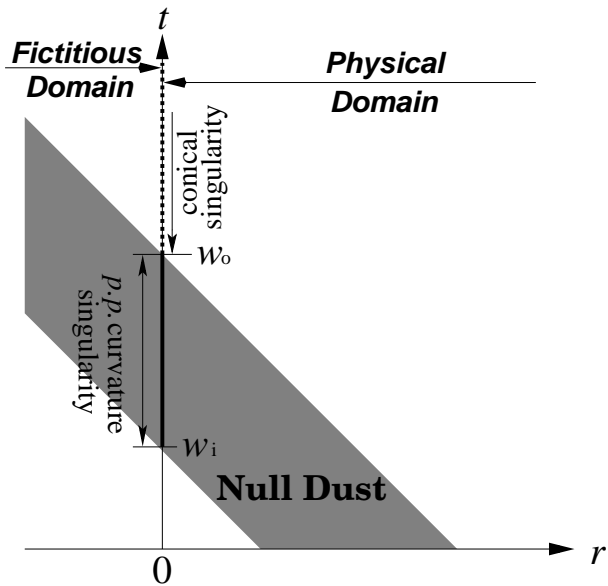}}}}
\caption{
Schematic diagram of the Morgan's null dust solution. 
}
\label{fg:null}
\end{figure}

It should be noted that all of the variables, $\gamma$, $\psi$, 
$R$ and $D$, are finite even at the 
spacetime singularity in the domain $0\leq r<+\infty$. 
Thus the domain for these variables can be extended from 
$0\leq r<\infty$ to $-\infty<r<\infty$. 
We may call the domain $0\leq r <\infty$ the physical domain, while we 
call the domain $-\infty<r<0$ the fictitious domain.  
In this extended domain, the condensation of the null dust into the spacetime 
singularity is regarded as a removal of the null dust 
from the physical domain to the fictitious one. 
For $t\leq w_{\rm i}$, whole of the null dust exists in the physical domain. 
At time $t=w_{\rm i}$, the inner surface of the hollow cylinder 
reaches the symmetry axis $r=0$. Whole of the cylinder 
enters into the fictitious domain by $t=w_{\rm o}$ (see Fig.\ref{fg:null}). 
The introduction of the fictitious domain is useful for  
constructing numerical solutions which describe the collapse of an imploding 
hollow cylinder composed of dust, in the next section. 

It is worthwhile to note again that, as mentioned in $\S$1, we 
have implicitly imposed some boundary condition at the naked singularity 
through the requirement for realizing the chronological future of the spacetime 
singularity as smooth as possible. 

\section{Imploding dust}

In this section, we consider the dust whose stress-energy tensor 
takes the same form as the null dust, 
\begin{equation} 
T^{ab}=\rho u^a u^b,
\end{equation}
but here $u^a$ is the unit timelike vector field whose integration 
curves are the world lines of dust particles. Here we write 
the components of $u^a$ in the form 
\begin{equation}
u^\mu=\frac{e^{-\gamma+\psi}}{\sqrt{1-v^2}}\left(1,v,0,0\right), 
\end{equation}
and we introduce a conserved density $D$ defined by
\begin{equation}
D=\sqrt{-g}\rho u^t=\frac{R e^{\gamma-\psi}\rho}{\sqrt{1-v^2}}~.\label{eq:dust-D-def}
\end{equation}
Note that this variable $D$ is equivalent to $D$ introduced 
in the preceding section. We also assume that $\rho$ is non-negative so that 
all of the reasonable energy conditions are satisfied. 
As in the case of the null dust, we assume that $D$ has a compact support in 
$r$-domain so that the dust constitutes a hollow cylinder. 

\subsection{The Einstein equations}

Since, as mentioned, the metric variable $R$ should vanish at $r=0$ before the 
singularity formation, we write it in the form, 
\begin{equation}
R=r\beta. 
\end{equation}
Then the Einstein equations become
\begin{eqnarray}
&&\gamma'=
\left\{\beta^2+2r\beta\beta'+r^2({\beta'}^{2}-{\dot \beta}^{2})\right\}^{-1}
\biggl[
r\beta\left(\beta+r\beta'\right)\left({\dot \psi}^{2}+{\psi'}^{2}\right)
-2r^2\beta\dot{\beta}{\dot \psi}\psi' \nonumber \\
&&~~~~+2\beta\beta'+r\left(2{\beta'}^2+\beta\beta''-\dot{\beta}^2\right)
+r^2\left(\beta'\beta''-\dot{\beta}\dot{\beta}'\right)\nonumber \\ 
&&~~~~+\frac{8\pi Ge^{\gamma-\psi}D}{\sqrt{1-v^2}}
\left(\beta+r\beta'+r\dot{\beta}v\right)
\biggr], 
\label{eq:constraint-1} \\
&&{\dot \gamma}=-
\left\{\beta^2+2r\beta\beta'+r^2({\beta'}^{2}-{\dot \beta}^{2})\right\}^{-1}
\biggl[
r^2\beta\dot{\beta}\left({\dot \psi}^{2}+{\psi'}^{2}\right)
-2r\beta\left(\beta+r\beta'\right){\dot \psi}\psi' \nonumber \\
&&~~~~-\beta\dot{\beta}+r\left(\dot{\beta}\beta'-\beta\dot{\beta}'\right)
+r^2\left(\dot{\beta}\beta''-\beta'\dot{\beta}'\right)\nonumber \\
&&~~~~+\frac{8\pi Ge^{\gamma-\psi}D}{\sqrt{1-v^2}}
\left\{r\dot{\beta}+\left(\beta+r\beta'\right)v\right\}
\biggr], 
\label{eq:constraint-2}\\
&&{\ddot \gamma}-\gamma''={\psi'}^{2}-{\dot \psi}^{2}, \label{eq:gamma-evol}\\
&&{\ddot \beta}-\beta''-{2\over r}\beta'
={8\pi G\over r}e^{\gamma-\psi}D\sqrt{1-v^2}, \label{eq:beta-evol}\\
&&{\ddot \psi}+{{\dot \beta}\over \beta}{\dot \psi}-\psi''
-{1\over r}\left(1+r{\beta'\over \beta}\right)\psi' 
={4\pi G\over r\beta}e^{\gamma-\psi}D\sqrt{1-v^2}.
\label{eq:psi-evol}
\end{eqnarray}
The equations of motion for dust $\nabla_a T^{ab}=0$ lead 
\begin{eqnarray}
\dot{D}+\left(vD\right)'&=&0, \label{eq:D-evol}\\
{dv\over dt}=\dot{v}+vv'&=&(1-v^2)\left\{v\left(\dot{\psi}-\dot{\gamma}\right)
+\psi'-\gamma'\right\}.
\label{eq:V-evol}
\end{eqnarray}
Equation (\ref{eq:D-evol}) represents the mass conservation, 
whereas Eq.(\ref{eq:V-evol}) is the geodesic equation. 

In contrast to the null dust case, it seems to be impossible to obtain solutions 
for the above equations analytically. Thus we invoke numerical simulations to 
study this system.

\subsection{Boundary condition}

It is the primary purpose of this paper to numerically construct solutions 
for the collapse of the dust similar to the Morgan's solution reviewed in the 
preceding section. When a hollow cylinder composed of dust collapses 
to the symmetry axis, $D$ has non-vanishing values at the symmetry axis. It is easily 
seen from Eq.(\ref{eq:dust-D-def}) that the rest mass density $\rho$ diverges 
at $r=0$ if $D$ does not vanish there, as long as $R=0$ at $r=0$, i.e., $\beta$ is 
finite there: as will be shown later, this is true in the present case. 
Thus, when the hollow cylinder of dust collapses to the symmetry 
axis, the spacetime singularity will form there. 

In the null dust case, $\beta=1$ and $\psi=0$ are the solutions for 
the Einstein equations, 
whereas these are not in the case of the dust. The motion of the 
dust disturbs $\beta$ and $\psi$. Thus, 
it is non-trivial whether $\beta$ and $\psi$ are still everywhere finite  
after the spacetime singularity appears at the symmetry axis. 
In the paper I, the present authors have studied the same subject, i.e., 
the collapse of a hollow cylinder composed of dust. By imposing 
appropriate boundary conditions 
on the metric and matter variables, $\beta$, $\gamma$, $\psi$, $D$ and $v$, 
at the spacetime singularity, the present authors constructed 
numerical solutions for these variables which are 
everywhere finite and continuous even after the formation of the 
spacetime singularity.  At that time, the 
present authors imposed {\it the going through boundary condition} on 
the motion of the dust: 
all of the dust particles collapsed to the spacetime singularity again expand 
with the same speed as their collapsing speed when they reach there. 

In the present case, the same boundary condition for the metric variables, 
$\beta$, $\gamma$ and $\psi$, at the spacetime singularity are also available, 
and we can construct $C^{2-}$ solutions for them. The chronological future 
of the spacetime singularity is realized and thus the resultant spacetime singularity 
is naked. 
By contrast, for the matter variables $D$ and $v$, we impose different boundary 
conditions from those assumed in the paper I. 
We set the boundary conditions on $D$ and $v$ so that the similar situation 
shown in the preceding section is realized i.e, the dust collapsed to the symmetry 
axis settles down there; we call this boundary condition {\it the settling down 
boundary condition}. 

First, we show the boundary conditions for the metric 
variables at the symmetry axis $r=0$. From Eq.(\ref{eq:beta-evol}), we have
\begin{equation}
\beta'=-4\pi Ge^{\gamma-\psi}D\sqrt{1-v^2} + r(\ddot{\beta}-\beta'').
\end{equation}
The above equation gives a Neumann boundary condition on $\beta$ at $r=0$ as
\begin{equation}
\beta'=-4\pi Ge^{\gamma-\psi}D\sqrt{1-v^2}~\bigr|_{r=0}. \label{eq:beta-bc}
\end{equation}
By the same procedure, we obtain 
the Neumann boundary conditions on $\psi$ and $\gamma$ 
at $r=0$ from Eqs. (\ref{eq:constraint-1}) and (\ref{eq:psi-evol}): 
\begin{eqnarray}
\gamma'&=&{8\pi Ge^{\gamma-\psi}Dv^2 \over \beta\sqrt{1-v^2}}\biggr|_{r=0},
\label{eq:gamma-bc} \\
\psi'&=&-{4\pi G \over \beta}e^{\gamma-\psi}D\sqrt{1-v^2}~\bigr|_{r=0}, 
\label{eq:psi-bc}
\end{eqnarray}
where we have used Eq.(\ref{eq:beta-bc}) to derive Eq.(\ref{eq:gamma-bc}).
These boundary conditions guarantee the finiteness of $\beta$, $\gamma$, $\psi$ 
and their derivatives with respect to $r$ as long as $D$ is finite and $v^2<1$. 
In other words, the spacetime singularity formed at the symmetry axis 
$r=0$ causes at most regular singular points in the Einstein equations 
as long as $D$ is finite and $v^2<1$. 
If $\dot{\gamma}$, $\gamma'$, $\dot{\psi}$ 
and $\psi'$ do not diverge, the right hand side of Eq.(\ref{eq:V-evol}) is 
also finite. Therefore, if initial data is appropriate, $D$ will 
remain finite and $v^2$ will be always less than unity 
even at the spacetime singularity. As a result, 
we will have finite and continuous solutions for all variables 
$\beta$, $\gamma$, $\psi$, $D$ and $v$ 
even at the spacetime singularity. 

Here we should note that non-vanishing derivatives of the 
metric variables with respect to $r$ 
at the symmetry axis imply the irregularity of the spacetime at the 
symmetry axis, even if the metric variables and thus the components 
of the metric tensor are $C^{2-}$ functions 
of $t$ and $r$. The functional regularity of the 
metric components is not equivalent to the spacetime regularity.\cite{NKMH:2007} 
It is seen from Eqs.(\ref{eq:beta-bc})--(\ref{eq:gamma-bc}) that 
if $D$ does not vanish at the symmetry axis $r=0$, 
the symmetry axis becomes singular. 

\begin{figure}
\centerline{\rotatebox{0}{\resizebox{8cm}{!}{\includegraphics{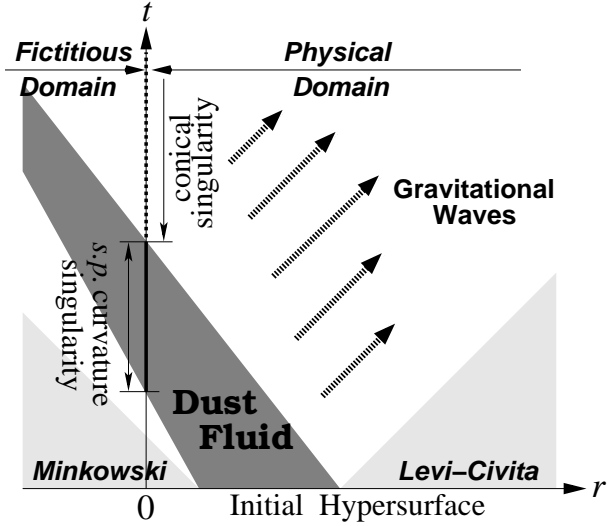}}}}
\caption{
Schematic diagram of the numerical domain. 
}
\label{fg:dust}
\end{figure}

As mentioned, in order to determine the boundary conditions 
on the matter variables $D$ and $v$ at the spacetime singularity, 
we refer the Morgan's null dust solution shown 
in $\S$3. We extend the domain $0<r<\infty$ to $-\infty<r< \infty$; 
as in the case of the null dust, we call the original 
domain {\it the physical domain} and the additional domain 
{\it the fictitious domain}. 
We assume appropriate $\gamma$ and $\psi$  
in this fictitious domain $r<0$; as in the paper I, 
we may assume $\gamma(t,r)=\gamma(t,-r)$ and  
$\psi(t,r)=\psi(t,-r)$. Then we solve  
Eqs.(\ref{eq:D-evol}) and (\ref{eq:V-evol}) for the fictitious domain  
as well as for the physical domain. 
We set the initial condition so that 
all of the collapsing dust will enter into the fictitious domain 
from the physical domain. When the dust passes through $r=0$, 
the symmetry axis becomes the spacetime singularity since 
$\rho$ and thus the Ricci scalar diverge there. 
In contrast to the null dust case, it is nontrivial whether 
there may remain the $\varphi$-angular deficit, 
after whole of the dust enters into the fictitious domain, or in other 
words, after whole of the dust is condensed into the spacetime singularity.  
By investigating Eq.(\ref{eq:constraint-2}), we find that 
the conically singular symmetry axis remains 
as a final product by the collapse of the hollow cylinder composed of dust. 
Equation (\ref{eq:constraint-2}) leads an evolution equation 
for $\gamma$ at $r=0$ as  
\begin{equation}
\left(\beta^{-1}e^\gamma\right)^{\cdot}=-\frac{8\pi G e^{2\gamma-\psi}Dv}{\beta^2\sqrt{1-v^2}}.
\end{equation}
Note that $\beta^{-1}e^\gamma$ should be unity on the regular axis. 
However, from the above equation, we can see that even if $\beta^{-1}e^\gamma$ 
is initially unity, it becomes larger than unity 
after the dust collapses to the symmetry axis $r=0$, since $v<0$ for collapsing dust. 
(To recognize the situation, see Fig.\ref{fg:dust}.) 

\section{Numerical Simulations}

\subsection{Initial Data and {\it C}-energy}

We set the initial conserved density and velocity field as 
\begin{eqnarray}
D&=&\frac{15\sigma}{32\pi w^5 l^5}
\left[r-l(1-w)\right]^2\left[r-l(1+w)\right]^2, \\
v&=&-\sqrt{1-\exp\left(-\frac{\mu}{r}\right)},
\end{eqnarray}
for $l(1-w)<r<l(1+w)$ and vanishes elsewhere, 
where $\sigma$ is the rest mass per unit Killing length in the direction 
with translational invariance, 
$l$ is a positive parameter to specify the mean-radius of the hollow dust cylinder, 
$w$ is a positive parameter smaller than unity, which specifies the thickness of 
the hollow dust cylinder, and $\mu$ is the parameter 
to control the gradient of the velocity field $v$. 

We set the initial data of metric variables, $\beta$, $\gamma$ and $\psi$ 
and their time derivatives in the following manner. 
We set $\beta=1$ and $\dot{\beta}=0$. Then the constraint 
equations (\ref{eq:constraint-1}) and (\ref{eq:constraint-2}) become
\begin{eqnarray}
\gamma'&=&r{\psi'}^2+\frac{8\pi Ge^{\gamma-\psi}D}{\sqrt{1-v^2}}, 
\label{eq:dgamma}\\
\dot{\gamma}&=&-\frac{8\pi Ge^{\gamma-\psi}Dv}{\sqrt{1-v^2}}.
\label{eq:dtgamma}
\end{eqnarray}
In order to obtain $\gamma$, 
we have got to integrate numerically Eq.(\ref{eq:dgamma}),  
whereas Eq.(\ref{eq:dtgamma}) gives directly the time derivative of $\gamma$. 
We are interested in the initial situation 
similar to the static configuration as well as possible although 
the initial imploding velocity does not vanish. 
Thus we set $\dot{\psi}=0$. 
In order to determine the initial data of $\psi$, we use 
Eq.(\ref{eq:psi-evol}) with $\ddot{\psi}=0$,
\begin{equation}
\psi''=-\frac{1}{r}\left(\psi'+4\pi Ge^{\gamma-\psi}D\sqrt{1-v^2}\right).
\label{eq:dpsi}
\end{equation}
We numerically integrate Eqs.(\ref{eq:dgamma}) and (\ref{eq:dpsi}) 
simultaneously outward from $r=0$ by imposing 
the boundary conditions $\psi|_{r=0}=0=\psi'|_{r=0}$ and $\gamma|_{r=0}=0$ 
which guarantee the regularity of the initial data. 

The vacuum region of the initial data obtained by the above procedure 
agrees with the Levi-Civita solution, 
\begin{equation}
\psi=-\kappa \ln r,
 ~~~~\gamma=\kappa^2 \ln r+\lambda~~~~{\rm and}~~~~\beta=1, \label{eq:LC}
\end{equation}
where $\kappa$, $\lambda$ are constant numbers which characterize this solution. 
Integrating Eq.(\ref{eq:dpsi}), we find that $\kappa$ vanishes for 
$r\leq l(1-w)$. Then from the regularity condition $\gamma|_{r=0}=0$, we have 
$\lambda=0$ for $0\leq r <l(1-w)$. For $r\geq l(1+w)$, we have 
\begin{equation}
\kappa=4\pi G\int_{l(1-w)}^{l(1+w)}dr e^{\gamma-\psi}D\sqrt{1-v^2}.
\end{equation}
Since $D$ is positive for $l(1-w)<r<l(1+w)$, $\kappa$ is 
positive in the domain $r\geq l(1+w)$. 

The $C$-energy $E_{\rm O}(t,r)$ proposed by Thorne 
is the quasi-local energy per unit Killing length in the direction with 
the translational invariance for the spacetime with 
whole cylinder symmetry.\cite{Thorne:1965} Its definition is given by 
\begin{equation}
E_{\rm O}(t,r)=\frac{1}{4G}\left[\gamma
-\frac{1}{2}\ln\left\{{\left(r\beta\right)'}^2
-(r\dot{\beta})^2\right\}\right].
\end{equation}
Substituting Eq.(\ref{eq:LC}) into the above equation, we have 
\begin{equation}
E_{\rm O}(t=0,r)={1\over4G}\left(\kappa^2\ln r+\lambda\right)  \label{eq:original}
\end{equation}
for the vacuum region, $0\leq r\leq l(1-w)$ or $r\geq l(1+w)$, of the present initial data. 
Since both of $\kappa$ and $\lambda$ vanish in the inside vacuum region, 
the $C$-energy vanishes for $0\leq r \leq l(1-w)$. 
The total $C$-energy is obtained by taking the limit of 
$r\rightarrow\infty$. We can easily see that the total 
$C$-energy of the present initial data is infinite 
irrespective of the values of $\lambda$ and $\kappa$. 
This means that the total $C$-energy is infinite irrespective of $D$ and $v$, 
but this fact is not so terrible. The similar situation 
is realized also in the Newtonian cylindrically symmetric system; 
the logarithmic divergence of the Newtonian gravitational potential 
at the spatial infinity necessarily leads infinite gravitational binding energy. 

In order to make the total energy per unit translational Killing length 
finite for the Levi-Civita spacetime, 
Thorne also proposed another definition of the $C$-energy as 
\begin{equation}
E_{\rm N}=\frac{1}{8G}\left(1-e^{-8GE_{\rm O}}\right)
=\frac{1}{8G}\left[1+e^{-2\gamma}\left\{(r\dot{\beta})^2
-{\left(r\beta\right)'}^2\right\}\right].
\end{equation}
Substituting Eq.(\ref{eq:LC}) into the above equation, we have  
\begin{equation}
E_{\rm N}={1\over8G}\left(1-e^{-2\lambda}r^{-2\kappa^2}\right)  \label{eq:LC-energy}
\end{equation}
By taking a limit of $r\rightarrow\infty$ in the above equation,   
we find that the total value of $E_{\rm N}$  
is equal to $1/8G$ irrespective of the values of $\lambda$ and $\kappa$. 
As shown by Hayward, $1/8G$ is the upper bound 
of total value of the new $C$-energy if the null energy condition is satisfied and 
if there is no singularity in the initial 
data\cite{Hayward:2000}. 

\subsection{Evolution}

\begin{figure}[htbp]
\centering
\includegraphics[scale=0.8,angle=0]{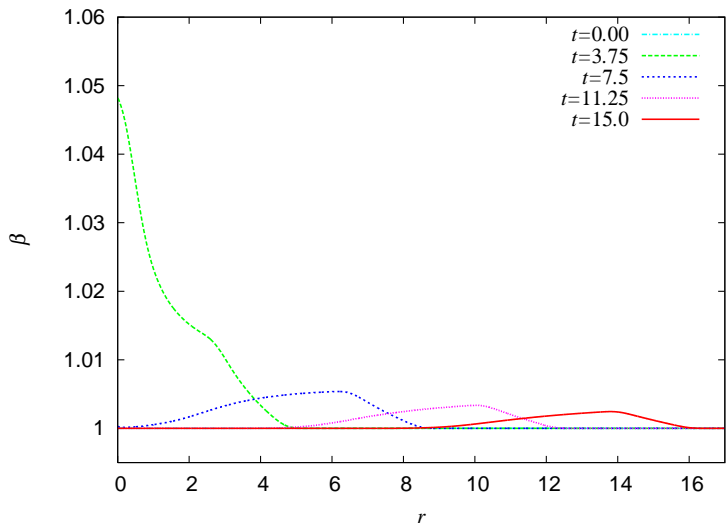}
\caption{
Several snapshots of $\beta$ in the physical domain $r\geq0$. The horizontal axis 
represents the radial coordinate $r$. 
}
\label{fg:beta}
\end{figure}

\begin{figure}[htbp]
\centering
\includegraphics[scale=0.8,angle=0]{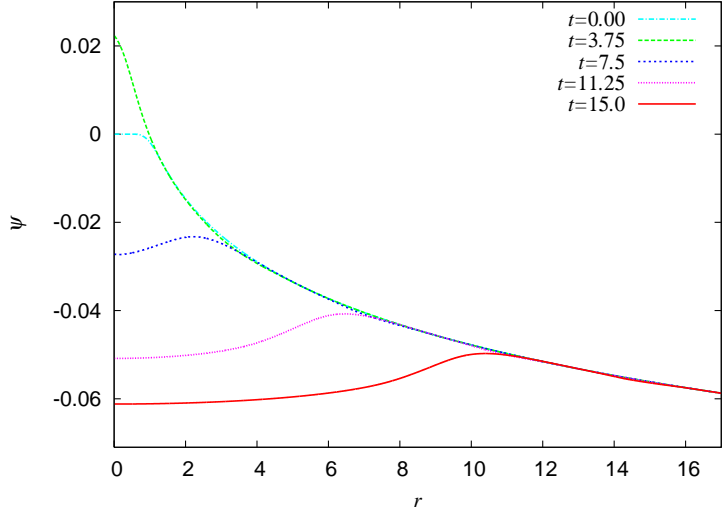}
\caption{
The same as Fig.\ref{fg:beta}, but for $\psi$. 
}
\label{fg:psi}
\end{figure}

\begin{figure}[htbp]
\centering
\includegraphics[scale=0.8,angle=0]{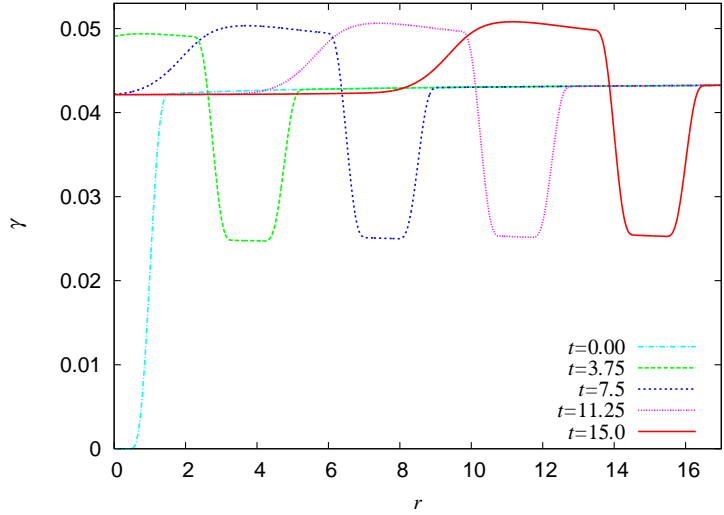}
\caption{
The same as Fig.\ref{fg:beta}, but for $\gamma$. 
}
\label{fg:gamma}
\end{figure}

\begin{figure}[htbp]
\centering
\includegraphics[scale=0.8,angle=0]{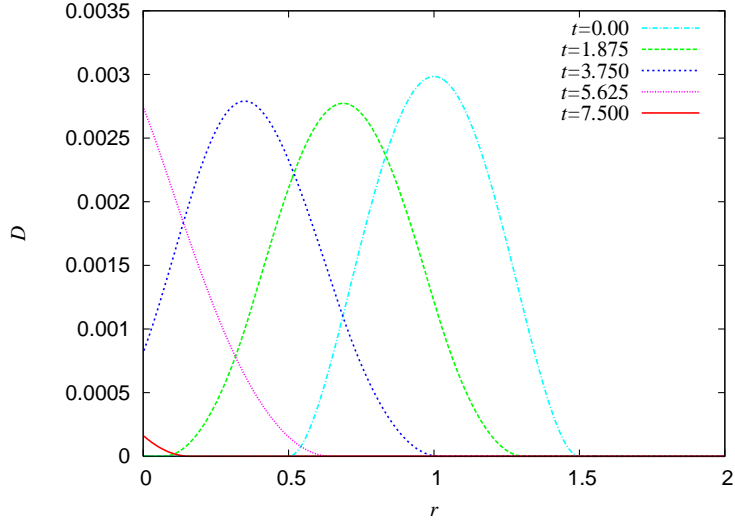}
\caption{
Several snapshots of the conserved rest mass density $D$ in the physical domain. 
The horizontal axis also represents the radial coordinate $r$. Note that 
the depicted range is different from those of Figs.\ref{fg:beta}--\ref{fg:gamma}
}
\label{fg:density}
\end{figure}

\begin{figure}[htbp]
\centering
\includegraphics[scale=0.8,angle=0]{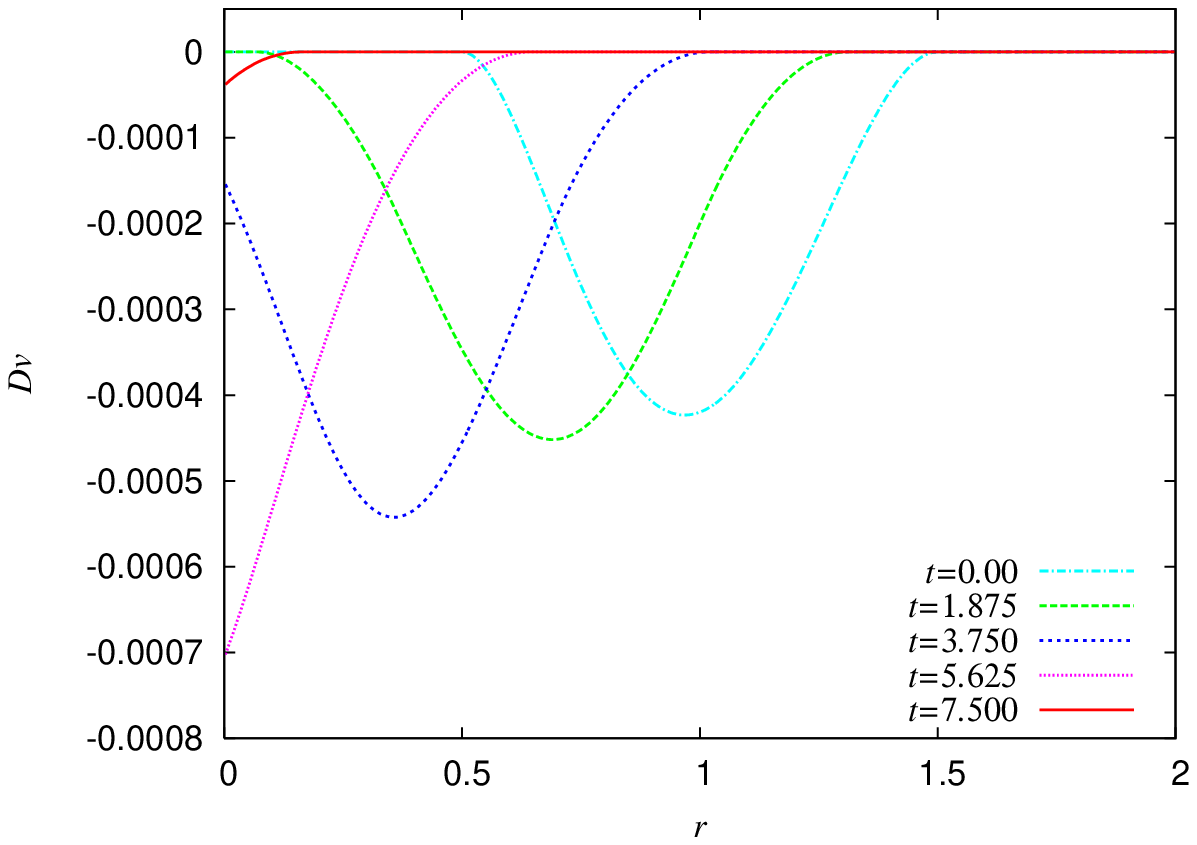}
\caption{
The same as Fig.\ref{fg:density}, but for the momentum density $Dv$. 
}
\label{fg:momentum}
\end{figure}

In order to study the dynamical behavior of the dust and spacetime geometry, 
we numerically integrate Eqs.(\ref{eq:gamma-evol})-(\ref{eq:V-evol}). 
We adopt the finite difference method and MacCormack scheme to solve 
the equations for the metric variables  
Eqs.(\ref{eq:gamma-evol})-(\ref{eq:psi-evol}). 
In order to solve the equations of motion for the dust (\ref{eq:D-evol}) 
and (\ref{eq:V-evol}), we adopt the method invented by Shapiro and Teukolsky; we 
follow the motion of large numbers of cylindrical mass shells which move along timelike 
geodesics and then construct the conserved rest mass density $D$ and 
the velocity field $v$ from their positions and 
velocities\cite{Shapiro:1991}. 
We show numerical solutions for $\beta$, $\psi$, $\gamma$,  
$D$ and $Dv$ with the parameters $l=1$, $w=0.5$, $\sigma=10^{-2}$ 
and $\mu=10^{-2}$ in Figs.\ref{fg:beta}--\ref{fg:momentum}. 
The numerically covered domain is $0\leq r\leq 15$. The numbers of 
the spatial grid points and the mass shells are $3\times 10^3$ and $300$, 
respectively. 

The constraint equations (\ref{eq:constraint-1}) 
and (\ref{eq:constraint-2}) are satisfied 
by the numerical solutions for Eqs.(\ref{eq:gamma-evol})-(\ref{eq:psi-evol}) 
only if these numerical solutions are good approximation of true solutions for 
the Einstein equations. Thus Eqs.(\ref{eq:constraint-1}) and 
(\ref{eq:constraint-2}) are used for monitoring the accuracy 
of numerical integrations of Eqs.(\ref{eq:gamma-evol})-(\ref{eq:psi-evol}).  
The relative errors estimated by the constraint equations 
are less than $10^{-3}$ in the numerical data shown in these figures. 


The inner surface of the hollow cylinder 
reaches the symmetry axis $r=0$ at $t\simeq 3.0$. 
As mentioned, when the dust reaches the symmetry axis 
$r=0$, the rest mass density $\rho$ and thus the Ricci scalar blow up there, 
since $\beta$ is finite there. 
This implies that the {\it s.p.} curvature singularity\cite{Hawking-Ellis} 
forms there. Since, as can be seen from these figures, all of the metric variables 
$\beta$, $\gamma$ and $\psi$ are everywhere 
finite and continuous, there is the chronological future of this singularity. 
Therefore a naked singularity forms in the spacetime constructed 
by this numerical simulation. 

\section{Asymptotic Behavior and Physical Implication}

The numerical simulations showed that a ripple in the metric variable $\beta$ 
propagates to infinity and $\beta$ asymptotically approaches to unity. 
By contrast, $\psi$ and $\gamma$ show different behaviors.  
After whole of the dust cylinder is condensed into the spacetime 
singularity, any characteristic scales 
disappear in this system. Thus the self-similar behaviors 
are expected for the metric variables $\psi$ and $\gamma$. 
In order to obtain asymptotic solutions for $\psi$ and $\gamma$ analytically, 
we introduce a variable defined by 
\begin{equation}
\xi=\frac{t-t_{\rm s}}{r},
\end{equation} 
where $t_{\rm s}$ is a constant. Then the metric variables $\psi$ and $\gamma$ are 
expected to asymptotically 
depend on only $\xi$, but, as shown below, this is not true. 
Assuming that $\beta=1$ and 
$\psi$ depends on only $\xi$, Eq.(\ref{eq:psi-evol}) becomes 
\begin{equation}
(1-\xi^2)\frac{d^2\psi}{d\xi^2}-\xi\frac{d\psi}{d\xi}=0. \label{eq:ss-psi-eq}
\end{equation}
If we also assume that $\gamma$ depends on only $\xi$, we have, 
from Eqs.(\ref{eq:constraint-1}) and (\ref{eq:constraint-2}), 
\begin{eqnarray}
\frac{d\gamma}{d\xi}&=&-\frac{1}{\xi}(1+\xi^2)\left(\frac{d\psi}{d\xi}\right)^2, \\ 
\frac{d\gamma}{d\xi}&=&-2\xi\left(\frac{d\psi}{d\xi}\right)^2.
\end{eqnarray}
The above two equations lead $d\psi/d\xi=0=d\gamma/d\xi$; 
there are no non-trivial vacuum solutions for $\psi$ and $\gamma$ which 
depend on only $\xi$. 

We construct asymptotic solutions in the following manner. 
First, we solve Eq.(\ref{eq:ss-psi-eq}). We have for $|\xi|\leq 1$
\begin{equation}
\psi=-\kappa_{\rm s}\sin^{-1}\xi+\psi_{\rm s},
\label{eq:inapp-psi}
\end{equation} 
and for $|\xi|>1$
\begin{equation}
\psi=-\kappa_{\rm s}\ln\biggl|\xi+\sqrt{\xi^2-1}\Biggr|+\psi_{\rm s},
\label{eq:true-ss-psi}
\end{equation}
where $\kappa_{\rm s}$ and $\psi_{\rm s}$ are integration constants. 
Since we are interested in the late time asymptotic behavior, the solution 
(\ref{eq:inapp-psi}) is inappropriate. The solution (\ref{eq:true-ss-psi}) 
diverges logarithmically at $r=0$ and thus  
this solution itself also is not what we need here. However, since the evolution 
equation for $\psi$ is linear, the solution which is finite at $r=0$ is obtained 
by superposing the Levi-Civita solution on the solution (\ref{eq:true-ss-psi}):  
\begin{equation}
\psi
=-\kappa_{\rm s}\ln\biggl|\xi+\sqrt{\xi^2-1}\Biggr|-\kappa_{\rm s}\ln r+\psi_{\rm s}
=-\kappa_{\rm s}\ln\biggl|t-t_{\rm s}
+\sqrt{(t-t_{\rm s})^2-r^2}\biggr|+\psi_{\rm s}. \label{eq:ss-psi}
\end{equation}
Next we assume that $\gamma$ depends on only $\xi$. 
Substituting Eq.(\ref{eq:ss-psi}) into Eqs.(\ref{eq:constraint-1}) 
and (\ref{eq:constraint-2}), we have an identical equation
\begin{equation}
\frac{d\gamma}{d\xi}=\frac{2\kappa_{\rm s}^2\left(\sqrt{\xi^2-1}-\xi\right)}
{\xi^2-1}.
\end{equation}
We can easily integrate the above equation and obtain
\begin{equation}
\gamma=2\kappa_{\rm s}^2
\ln\Biggl|\frac{\xi+\sqrt{\xi^2-1}}{2\sqrt{\xi^2-1}}\Biggr|
+\lambda_{\rm s}, \label{eq:ss-gamma}
\end{equation}
where $\lambda_{\rm s}$ is an integration constant. Eqs.(\ref{eq:ss-psi}) 
and (\ref{eq:ss-gamma}) are the solutions that we need. 
The Kretschimann invariant of this spacetime is given by
\begin{eqnarray}
R^{\mu\nu\rho\sigma}R_{\mu\nu\rho\sigma}
&=&2^{4+8\kappa_{\rm s}^2}\kappa_{\rm s}^2(1+2\kappa_{\rm s})^2
e^{4(\psi_{\rm s}-\lambda_{\rm s})}(\tau+\sqrt{\tau^2-r^2})^{-2(4\kappa_{\rm s}^2+2\kappa_{\rm s}+1)}
(\tau^2-r^2)^{4\kappa_{\rm s}^2-3/2} \nonumber \\
&\times&\left[
\left(1-\kappa_{\rm s}-2\kappa_{\rm s}^2\right)\tau
+\left(2+\kappa_{\rm s}+2\kappa_{\rm s}^2\right)\sqrt{\tau^2-r^2}\right],
\end{eqnarray}
where $\tau=t-t_{\rm s}$. At $r=|t-t_{\rm s}|$, 
the Kretschimann invariant diverges in the case 
of $\kappa_{\rm s}^2 < 3/8$. All of the numerical solutions presented 
in this paper satisfy $\kappa_{\rm s}^2<3/8$. Rigorous derivation of this solution from a 
view point of self-similarity is given by two of the present authors, TH and KN, and 
their collaborator Nolan.\cite{HNN}.

\begin{figure}[htbp]
\centering
\includegraphics[scale=0.8,angle=0]{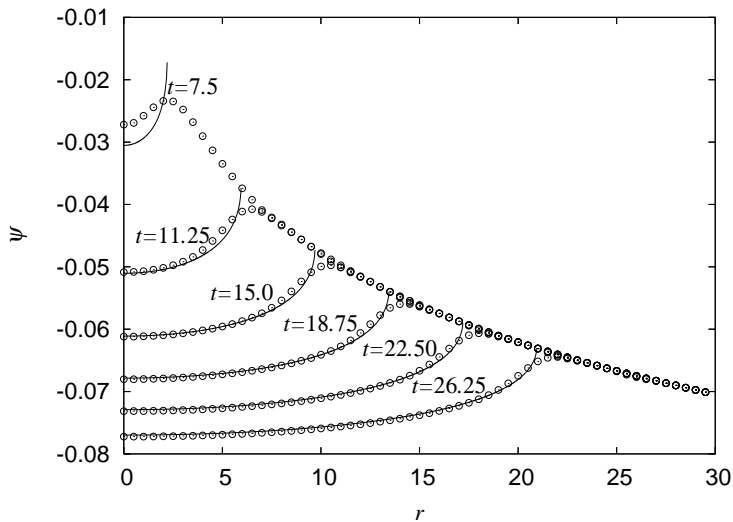}
\caption{Several snapshots of the metric variable $\psi$ are depicted; white circles 
represent the numerical values, whereas the solid curves represent 
the analytic solution (\ref{eq:ss-psi}). Note the plotted range.  
}
\label{fg:ss-psi}
\end{figure}
\begin{figure}[htbp]
\centering
\includegraphics[scale=0.8,angle=0]{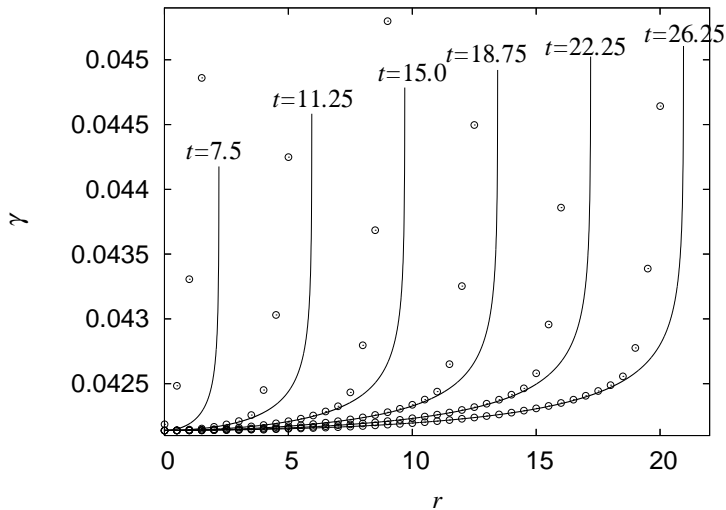}
\caption{The same as Fig.\ref{fg:ss-psi}, but for $\gamma$. The plotted range for 
each snapshot is restricted to the neighborhood of the symmetry axis $r=0$. 
}
\label{fg:ss-gamma}
\end{figure}

The solutions (\ref{eq:ss-psi}) and (\ref{eq:ss-gamma}) are 
depicted in Figs.\ref{fg:ss-psi} and \ref{fg:ss-gamma} together with  
numerical solutions; the numerically covered domain is 
$0\leq r \leq 30$, and the numbers of the grid points and the mass shells 
are $6\times 10^3$ and $300$, respectively. We set the parameters 
$\kappa_{\rm s}$ and $\lambda_{\rm s}$ 
to be equal to the numerical values $\kappa=2.06\times 10^{-2}$ and 
$\gamma|_{t=30}=4.21\times 10^{-2}$, respectively. 
Then we set $t_{\rm s}=5.30$ so that the 
analytic solutions agree well with numerical data. 
The analytic solution is available only for 
$0\leq r < |t-t_{\rm s}|$, and hence we have plotted the data for this domain. 
It is seen from these 
figures that the numerical solutions asymptotically approach to these 
analytic solutions in the neighborhood of the symmetry axis $r=0$. 
Therefore, even if we do not invoke 
long time numerical simulations, we can know the asymptotic 
behavior through these analytic solutions; 
$\psi$ monotonically decreases, whereas $\gamma$ approaches to $\lambda_{\rm s}$. 
Since the ripples in $\beta$ propagate outward in the manner 
$\beta\sim 1+f(t-r)/r$ in late time, where $f(x)$ is a function of 
compact support, the $C$-energy at any finite radial coordinate $r$ 
has a following limit
\begin{equation}
\lim_{t\rightarrow\infty}E_{\rm N}=\frac{1}{8G}(1-e^{-2\lambda_{\rm s}}).
\end{equation} 
Since the final $E_{\rm N}$ is constant, the 
$C$-energy concentrates to the symmetry axis in the final configuration. 

The distance from the symmetry axis  
to a point labeled by a non-vanishing radial coordinate $r$ becomes larger 
as time goes on, because $\psi\rightarrow-\infty$ and 
$\gamma\rightarrow \lambda_{\rm s}$ for $t\rightarrow\infty$. 
Further the Riemann tensor $R^{\mu\nu}{}_{\rho\sigma}$ 
behaves as $t^{-2(1+\kappa_{\rm s})}$ asymptotically 
at any radial coordinate $r$ (see Appendix). Thus the final spacetime is 
flat except at the symmetry axis $r=0$ which is conically 
singular. Due to the settling down boundary condition, the remnant of the
collapse of an imploding hollow cylinder of dust 
is the same as that of the null dust. 

\begin{figure}[htbp]
\centering
\includegraphics[scale=0.8,angle=0]{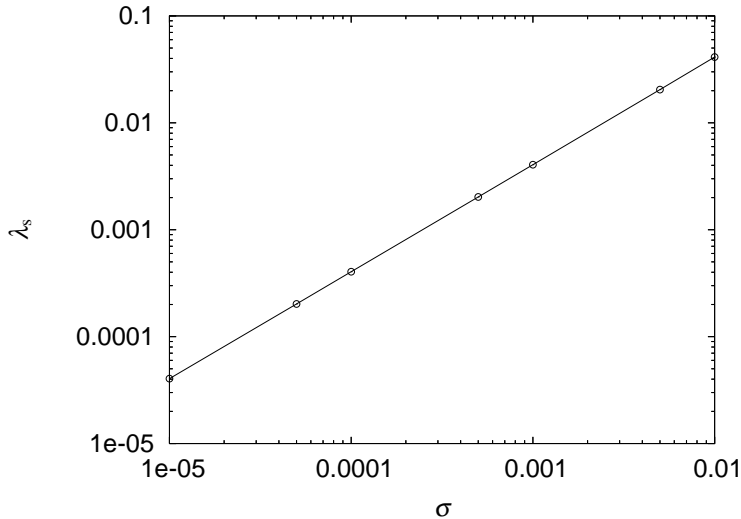}
\caption{
The relation between $\lambda_{\rm s}$ and $\sigma$ is depicted. 
}
\label{fg:sigma_lambda}
\end{figure}

The total energy per unit translational Killing length decreases by the emission of 
gravitational radiation. The total value of the 
new $C$-energy $E_{\rm N}$ is initially equal to $1/8G$, while 
it finally becomes $(1-e^{-2\lambda_{\rm s}})/8G$. 
The energy $e^{-2\lambda_{\rm s}}/8G$ has been released by the gravitational radiation. 
The numerical results imply that $\lambda_{\rm s}\simeq 4\sigma/G$ 
for $10^{-5}\leq\sigma/G \leq 10^{-2}$ (see Fig.\ref{fg:sigma_lambda}). 
Accordingly the ratio of the emitted energy to the initial one 
$\varepsilon=e^{-2\lambda_{\rm s}}$ depends on $\sigma$ in the manner
\begin{equation}
\varepsilon\simeq e^{-8\sigma/G}.  
\label{eq:efficiency}
\end{equation}
The lighter the weight of the dust cylinder is, the larger the 
efficiency $\varepsilon$ becomes. In zero-mass limit, the efficiency 
$\varepsilon$ becomes unity. This seems to be paradoxical, but we should note that 
it is a non-trivial matter how to relate the present results with the asymptotically 
flat cases, e.g., the situation treated by 
Shapiro and Teukolsky\cite{Shapiro:1991}. 
From the point of view of the original $C$-energy $E_{\rm O}$, 
the infinite amount of energy is released by the collapse of the dust cylinder; 
we can easily see that total value of the original $C$-energy finally becomes  
\begin{equation}
\lim_{t\rightarrow\infty}E_{\rm O}=\frac{\lambda_{\rm s}}{4G}
\end{equation}
for any radial coordinate $r$, whereas it is initially infinite. 

\section{Summary and Discussion}

We constructed numerical solutions for the Einstein equations, which 
describe the collapse of an imploding hollow cylinder composed of dust 
with a requirement that the dust particles stay the symmetry axis 
after these reach there. A spacetime singularity forms at its symmetry axis. 
Although the rest mass density and the curvature polynomials 
blow up at the spacetime singularity, this spacetime singularity causes 
at most regular singular points in Einstein equations. 
Thus if appropriate boundary conditions  
are imposed, components of the metric tensor are everywhere finite. 
Then the causal future of the spacetime singularity 
exists and the resultant spacetime singularity is naked. 

We also obtained an analytic solution which asymptotically well agrees with 
our numerical solutions. This asymptotic solution  
reveals that infinite amount of energy per unit translational Killing 
length is released to infinity by the gravitational radiation, 
and a conical singularity remains at the symmetry axis 
as a final product. Strictly speaking, this is a counterexample 
of the second type for the weak cosmic censorship conjecture. 
However, since this naked singularity is merely conical, 
it is not so serious.

It might be a surprising fact for some readers 
that the remnant naked singularity formed by gravitational 
collapse of a dust cylinder is weak. 
In Newtonian theory of gravity, collapse of a dust cylinder finally produces a 
gravitational potential which logarithmically diverges at its symmetry axis, and thus 
the singularity is strong in Tipler's sense. 
The reason of this difference between Newtonian gravity and 
relativity is as follows. 
The Newtonian gravitational potential $\Phi$ produced by a dust cylinder 
of radius $r=\ell$ is 
\begin{equation}
\Phi=-2G\sigma\ln \left(\frac{r}{r_{\rm c}}\right)~~~~~~{\rm for}~~r\geq\ell,
\end{equation}
where $r_{\rm c}$ is an integration constant, and 
$\sigma$ is the mass per unit length,  
\begin{equation}
\sigma=2\pi \int_0^\ell D(t,x)dx.
\end{equation}
Usually, we assume that $\sigma$ is conserved in the framework of Newtonian 
gravity. Then $\Phi$ logarithmically 
diverges at $r=0$ when the dust cylinder becomes infinitesimally thin, i.e., $\ell=0$. 
By contrast, the present prescription does not guarantee the constancy 
of $\sigma$, and require that $\sigma$ vanishes finally. 
Thus, if the present prescription is adopted in the framework of Newtonian 
gravity, the Newtonian gravitational potential finally 
vanishes. This prescription seems to be unphysical from a point of view 
of the mass conservation in Newtonian theory. However, it is not so 
from a point of view of the $C$-energy conservation in general relativity. 
Although the tidal force finally vanishes except on the symmetry axis, 
the $C$-energy is condensed on the symmetry axis, and produces a conical singularity there, 
in the framework of general relativity. The present prescription 
requires that the equation of state changes from dust to something other than dust, 
and the Newtonian approximation is not applicable to this changed equation of state. 

Here it might be useful to compare the present analysis with that of paper I. 
In paper I, we required that the dust particles pass through 
the symmetry axis after these reach there. In the resultant spacetime, 
almost all of geodesics are complete and thus this may not be a counterexample 
for the extended weak cosmic censorship conjecture, as mentioned in Sec.1. 
This example revealed that the gravity produced by the collapsed dust cylinder 
is too weak to confine the collisionless particles to the symmetry axis. 
In contrast, in the present prescription, matter 
is confined to the symmetry axis, but it is due to not the gravity, but 
the change of interactions between particles, or the change of 
the equation of state.  It should be noted that 
this change in the equation of state 
is a result by requiring the smoothest behaviors of the 
metric components like as the Morgan's null dust solution. 

Finally, we note that, in order to complete an analysis to see whether 
a numerical model like as that of Shapiro and Teukolsky is a counterexample for 
the cosmic censorship conjecture, we need to specify 
the boundary conditions at the ``spacetime singularity''. If we do not so, we can not know 
whether the ``singularity'' is really singular, or 
whether the cosmic censorship conjecture holds in a real singularity case.  
For this purpose, we need the knowledge of the global structure of the spacetime. In this sense, 
the analyses of paper I and the present paper are the first step of the attempt toward 
the numerical study of the weak cosmic censorship conjecture. 

\section*{Acknowledgments}

We are grateful to H.~Ishihara and colleagues in the astrophysics and 
gravity group of Osaka City University for their useful and helpful 
discussion and criticism. 
This work is supported by the Grant-in-Aid for Scientific 
Research (No.16540264). 

\appendix
\section{Riemann tensor of self-similar gravitational waves} 

The components of the Riemann tensor of the self-similar gravitational 
waves (\ref{eq:ss-psi}) and (\ref{eq:ss-gamma}) are given as follows:
\begin{eqnarray}
R^{tr}{}_{tr}&=&
-2^{4\kappa_{\rm s}^2}\kappa_{\rm s}(1+2\kappa_{\rm s})e^{2(\psi_{\rm s}-\lambda_{\rm s})}
(\tau^2-r^2)^{2\kappa_{\rm s}^2-1/2} 
\left(\tau+\sqrt{\tau^2-r^2}\right)^{-(4\kappa_{\rm s}^2+2\kappa_{\rm s}+1)} \nonumber \\
&=&R^{\varphi z}{}_{\varphi z}, \\
R^{r\varphi}{}_{r\varphi}&=&
2^{4\kappa_{\rm s}^2}\kappa_{\rm s}(1+2\kappa_{\rm s})e^{2(\psi_{\rm s}-\lambda_{\rm s})}
(\tau^2-r^2)^{2\kappa_{\rm s}^2-3/2} 
\left(\tau+\sqrt{\tau^2-r^2}\right)^{-(4\kappa_{\rm s}^2+2\kappa_{\rm s}+1)}
\nonumber \\
&\times&\left[(1-\kappa_{\rm s})\tau^2+(1+\kappa_{\rm s})\tau\sqrt{\tau^2-r^2}-\kappa_{\rm s}r^2\right] 
\nonumber\\
&=&R^{tz}{}_{tz}, \\
R^{rz}{}_{rz}&=&
2^{4\kappa_{\rm s}^2}\kappa_{\rm s}(1+2\kappa_{\rm s})e^{2(\psi_{\rm s}-\lambda_{\rm s})}
(\tau^2-r^2)^{2\kappa_{\rm s}^2-3/2} 
\left(\tau+\sqrt{\tau^2-r^2}\right)^{-(4\kappa_{\rm s}^2+2\kappa_{\rm s}+1)}
\nonumber \\
&\times&\left[\kappa_{\rm s}\tau^2-(1+\kappa_{\rm s})\tau\sqrt{\tau^2-r^2}-(1+\kappa_{\rm s})r^2\right]
\nonumber \\
&=&R^{t\varphi}{}_{t\varphi}, \\
R^{tz}{}_{rz}&=&
-2^{4\kappa_{\rm s}^2}\kappa_{\rm s}(1+2\kappa_{\rm s})e^{2(\psi_{\rm s}-\lambda_{\rm s})}
r(\tau^2-r^2)^{2\kappa_{\rm s}^2-3/2} 
\left(\tau+\sqrt{\tau^2-r^2}\right)^{-(4\kappa_{\rm s}^2+2\kappa_{\rm s}+1)}
\nonumber \\
&\times&\left[(1-2\kappa_{\rm s})\tau+(1+\kappa_{\rm s})\sqrt{\tau^2-r^2}\right] 
\nonumber \\
&=&-R^{t\varphi}{}_{r\varphi}. 
\end{eqnarray}

\end{document}